\documentclass[showpacs,twocolumn,floatfix]{revtex4}
\usepackage{graphicx}
\usepackage{dcolumn}
\usepackage{bm}
\usepackage{amssymb}
\usepackage{amsmath}
\usepackage{epsf}
\usepackage{subfigure}
\usepackage{epstopdf}%
\usepackage{amsfonts}
\usepackage{color}

\begin{document}

\title{Pulse propagation in decorated random chains}
\author{Upendra Harbola$^1$, Alexandre Rosas$^2$, Aldo H. Romero$^3$ and Katja Lindenberg$^1$}
\affiliation{${}^1$Department of Chemistry and Biochemistry, University of California San Diego, La
Jolla, California 92093-0340, USA.\\
${}^2$Departamento de F\'{\i}sica, CCEN, Universidade Federal da Para\'{\i}ba, Caixa Postal
5008, 58059-900, Jo\~ao Pessoa, Brazil.\\
${}^3$Cinvestav-Quer\'etaro, Libramiento Norponiente 200, 76230, Fracc. Real de
Juriquilla, Quer\'etaro, Quer\'etaro, M\'exico\\}

\date{\today}

\begin{abstract}
We study pulse propagation in one-dimensional chains of spherical
granules decorated with small randomly-sized granules placed between bigger monodisperse ones. 
Such ``designer chains" are of interest in efforts to control the behavior of the pulse
so as to optimize its propagation or attentuation, depending on the desired application. 
We show that a recently proposed effective description of simple decorated chains
can be extended to predict pulse properties in chains decorated with small granules of randomly
chosen radii.  Furthermore, we also show that the binary collision approximation can again be
used to provide analytic results for this system.
\end{abstract}

\pacs{46.40.Cd,43.25.+y,45.70.-n,05.65.+b}

\maketitle 

\section{Introduction}
\label{introduction}

The study of pulse propagation in granular media has become a field
of intense research interest.  This is partly due to the fundamental importance of
understanding the associated nonlinear dynamics and partly due to its direct application to our
day-to-day lives. Any system in which interactions among its discrete constituents 
are solely described by their macroscopic geometrical shapes and elasticity properties
rather than by their microscopic (atomic or molecular) nature can be classified as granular system. 
Among granular systems, a class that draws special attention is that of so-called dry granular
systems. A defining property of these systems is that the intergranule interaction
potentials are always positive i.e., they are purely mutually repulsive.  Furthermore, they
are only nonzero as long as the two bodies are in physical contact. 
This peculiarity, together with the spatially discrete nature of granular systems,
gives rise to a host of interesting phenomena. Depending on parameter values,
granular systems can express liquid-like, solid-like (glasses), or gas-like
properties~\cite{KadanoffRMP99,SilbertPRL05}.

It is well known that an initial kinetic energy impulse imparted to an edge granule of a
chain in the absence of precompression can result in
solitary waves propagating through the medium~\cite{nesterenko,nesterenko-1}. 
In the recent past, pulse propagation in one-dimensional (1D) chains of granules has
been studied extensively both theoretically and
experimentally~\cite{nesterenko-book,application,alexandremono,alexandre,jean,wangPRE07,rosasPRL07,rosasPRE08,sen-review,costePRE97,nesterenkoJP94,
sokolowEPL07,wu02,robertPRE05,sokolowAPL05,senPHYA01,meloPRE06,jobGM07,senJAP09}.
These studies are in part inspired by a number of
practical applications, e.g., in the design of shock absorbers~\cite{DarioPRL06,HongPRL05},
sound scramblers~\cite{VitaliPRL05,DaraioPRE05} and
actuating devices~\cite{KhatriSIMP08}.

Structural variations along a granular chain can significantly influence pulse 
propagation through the system.  These structural variations (or polydispersity) are frequently
introduced in a regular fashion such as in tapered chains, in which the size and/or mass of
successive granules systematically decreases or increases. A detailed study of these effects
has been carried out
numerically~\cite{senJAP09,wangPRE07,sen-review,robertPRL06,wu02,robertPRE05,sokolowAPL05,senPHYA01}.
In our own recent work~\cite{PRE09} we introduced a binary collision model to derive fairly
simple analytic results to describe pulse propagation in various 1D tapered chains. 
We showed that most of the essential properties that characterize pulse propagation in tapered
chains can be extremely accurately described in these systems using the binary collision model.

Recently we turned our attention to more complex chain configurations, specifically, to decorated
chains, that is, to chains in which large and small granules alternate in some regular fashion.
These chains can not be studied directly using any binary collision model, and the reason is quite
obvious: as a pulse propagates, the small granules rattle back and forth between their larger
neighbors and thus at least three granules rather than two are involved in elementary collision
events. However, we introduced an effective description~\cite{PRE2} whereby we represented decorated
chains by associated undecorated ones.  We tested this methodology on a variety of simple and tapered
decorated chains and showed that the effective undecorated chains reproduced
the behavior of pulse propagation in the original decorated ones in all cases provided the small
decorating granules were sufficiently small (see below).  This effective representation can then be
treated analytically using the binary collision approximation.  The resulting analytic expressions
allow us to explore regimes in which numerical algorithms may be unstable, and even regimes where the
pulse amplitude becomes so weak as to be close to the numerical noise.

All the studies described above have dealt with chains with regular configurations, be they
monodisperse or tapered, simple or decorated.  Here we introduce a new element, namely randomness.
It is interesting to study the effects of randomness because (1) most
granular systems in nature are not regular, and (2) randomness might be used as a control element in
manipulating pulse propagation properties in designed systems.  It therefore becomes important to 
understand effects of randomness on pulse propagation. Except for recent work by
Fraternali et al.~\cite{DaraioMAMS09}, we are not aware of any other work on 
pulse propagation in random granular chains. In this work we develop an understanding of some
effects of randomness on pulse propagation.  For this purpose, we extend our previous work on decorated
chains~\cite{PRE2} to incorporate effects of randomness. 

Our randomly decorated chain is constructed from a monodisperse array of large granules
separated pairwise by smaller granules whose size
is randomly picked from a pre-assigned distribution. For our effective
description~\cite{PRE2} to be valid, the radius of the
smaller granules must be no larger than $\sim 40\%$ of the larger granules.
This places a restriction on the upper cutoff of the size distribution of smaller granules.
We show that a randomly decorated chain can then be mapped onto an effective undecorated
chain with random masses and random interactions.
We find that for all properties studied herein, the behavior of 
our effective chain (obtained from the numerical solution of the equations of motion for this chain)
is in remarkable agreement with the behavior of the original decorated chain (obtained from the
numerical solution of its equations of motion).
In particular, we focus on three properties of the
pulse. First, we compute the pulse amplitude and its variation along the chain.
Second, we determine the average
speed of the pulse along the chain. Third, we calculate the distribution
of the times that the pulse takes  to reach the end of the chain.

We then go on to test our binary collision approximation applied to the effective chain.
We find that whereas the approximation
does not reproduce the pulse amplitude well (for reasons that we understand and for which we
suggest a possible remedy), the other two properties are extremely well predicted analytically by the model.
This provides a powerful tool to avoid costly numerical simulations.

In Sec.~\ref{model} we introduce the granular chain model in terms of rescaled
(dimensionless) variables.  In Sec.~\ref{decorated-chains} we introduce decorated chains and
present our effective description in terms of undecorated chains with renormalized interactions and masses.
The analysis here is a generalization of the work of Ref.~\cite{PRE2} to accommodate the random
variation (within limits) of the radii of the smaller granules. 
In this section we also exhibit the analytic results obtained by applying the binary
collision approximation to the effective chain.
Comparisons with numerical results are presented in Sec.~\ref{numerics}. In Sec.~\ref{conclusion}
we provide a summarizing closure.

\section{The model}
\label{model}

We consider chains of granules all made of the same material of density $\rho$.  
When neighboring granules collide, they repel each other according to
the power law potential
\begin{equation}
 \label{hertz-1}
 V=\frac{a}{n} r_k^\prime |y_k-y_{k+1}|^n.
\end{equation}
Here $y_k$ is the displacement of granule $k$ from its position at the beginning of the collision,
and $a$ is a constant determined by Young's modulus and Poisson's ratio~\cite{landau,hertz}.
The exponent $n$ is $5/2$
for spheres (Hertz potential), which we use throughout this paper in our explicit 
calculations~\cite{hertz}. We have defined 
\begin{equation}
  r_k^\prime = \left(\frac{2 R_k^\prime R_{k+1}^\prime}{R_k^\prime +R_{k+1}^\prime }\right)^{1/2},
\end{equation}
where $R_k^\prime$ is the principal
radius of curvature of the surface of granule $k$ at the point of contact.
When the granules do not overlap, the potential is zero.
The equation of motion for the $k$th granule is
\begin{eqnarray}
 \label{EOM-1}
 M_k\frac{d^2y_k}{d{\tau}^2}&=& {a}{r}_{k-1}^\prime(y_{k-1}-y_{k})^{n-1}\theta(y_{k-1}-y_k)\nonumber\\
&-& {a}{r}_{k}^\prime(y_k-y_{k+1})^{n-1}\theta(y_k-y_{k+1}),
\end{eqnarray}
where $M_k=(4/3)\pi \rho (R_k^\prime)^3$.  The Heaviside function
$\theta(y)$ ensures that the elastic interaction between grains is only nonzero if they are in contact.  
Initially the granules are placed along a line so that they
just touch their neighbors in their equilibrium positions (no
precompression), and
all but the leftmost particle are at rest. The initial velocity of
the leftmost particle ($k=1$) is $V_1$ (the impulse). 
We define the dimensionless quantity
\begin{equation}
\alpha\equiv \left[ \frac{M_1V_1^2}{a \left(R_1^\prime\right)^{n+1/2}}\right]
\end{equation}
and the rescaled quantities $x_k$, $t$, $m_k$, and $R_k$ via the relations
\begin{eqnarray}
y_k = R_1^\prime\alpha^{1/n} x_k, &\qquad&
\tau = \frac{R_1^\prime}{V_1} \alpha^{1/n} t, \nonumber\\
R_k^\prime = R_1^\prime R_k, &\qquad& M_k=M_1 m_k.
\end{eqnarray}
Equation~(\ref{EOM-1}) can then be rewritten as
\begin{eqnarray}
  m_k \ddot{x}_k &=&
 r_{k-1}(x_{k-1} - x_k)^{n-1} \theta (x_{k-1} -
x_k) \nonumber\\
&&-  r_k (x_k - x_{k+1})^{n-1} \theta (x_k -
x_{k+1}),
\label{eq:motion_rescaled}
\end{eqnarray}
where a dot denotes a derivative with respect to $t$, and
\begin{equation}
 \label{r-prime}
r_k= \left(\frac{2 R_kR_{k+1}}{R_k+R_{k+1}}\right)^{1/2}.
\end{equation}
The rescaled initial velocity is unity, i.e., $v_1(t=0)=1$. 
The velocity of the $k$-th granule in unscaled variables is simply $V_1$ times its velocity in the
scaled variables.

\section{Theoretical results: effective chain and binary collision approximation}
\label{decorated-chains}

\subsection{Effective chain}
\label{effective}

We consider a decorated chain with a small granule inserted between each pair of large granules.
The end granules are large, and the large granules are monodisperse. The sizes of the small
granules are random.  In order to obtain an effective description for the pulse dynamics in the
decorated chain, we follow the scheme
presented in Ref.~\cite{PRE2}. In that work we showed that the effective description could be
constructed using the outcome of
a detailed analysis of a chain of five granules.  This five-granule chain
provides the elements of all the granules in the actual long chain: large granules in the interior
of the chain, small granules in the interior of the chain, and large granules at the ends of the
chain.

Consider, then, a chain of five granules labeled from
$k-2$ to $k+2$ in unit steps, granules $k-1$ and $k+1$ being the small granules. The radius of the
large granules is $R_1\equiv R =1$ in rescaled variables, and those of the small granules is
$R_k=R_k^\prime/R_1^\prime$.
The dynamics of this chain of granules is governed by the set of equations,
\begin{eqnarray}
 \label{lg-1}
m_{k-2}\ddot{x}_{k-2} &=&  -r_{k-2}(x_{k-2} - x_{k-1})^{n-1} \theta (x_{k-2} -x_{k-1})\nonumber\\ 
m_{i}\ddot{x}_{i} &=&  r_{i-1}(x_{i-1} - x_{i})^{n-1} \theta (x_{i-1} -x_i)\nonumber\\
                         &-& r_{i}(x_{i} - x_{i+1})^{n-1} \theta (x_i -x_{i+1}), ~~ i=k, k\pm 1,\nonumber\\
m_{k+2}\ddot{x}_{k+2} &=&  r_{k+1}(x_{k+1} - x_{k+2})^{n-1} \theta (x_{k+1} -x_{k+2}). 
\end{eqnarray}
Since all the large granules are of the same size, it follows that $m_{k-2}=m_{k+1}=m_{k+2}=m$
($=1$ in the rescaled units), and 
we have $r_{k-1}=r_{k-2}$ and $r_{k}=r_{k+1}$, which just says that for a given small granule
the left and right sides are geometrically identical. Following Ref.~\cite{PRE2}, this chain can be
mapped onto an effective chain
of three large granules which we call ``left" ($l$), ``right" ($r$), and ``between" ($b$), with
renormalized masses $\mu_l$, $\mu_r$ and $\mu_b$ given by
\begin{eqnarray}
\label{renorm-mass}
\mu_l &=& m+\frac{1}{2}m_{k-1},~~~ \mu_r = m+\frac{1}{2}m_{k+1}\nonumber\\
\mu_b &=& m+\frac{1}{2}(m_{k-1}+m_{k+1}).  
\end{eqnarray}
The effective interaction between two neighboring large granules in this three-grain configuration
is 
\begin{eqnarray}
\label{effec-int}
V_{eff}(k,k+1) = \frac{r_k}{n2^{n-1}} (x_k-x_{k+1})^n,
\end{eqnarray}
where
\begin{eqnarray}
 \label{zetak}
r_k = \sqrt{\frac{2R_{k}}{1+R_{k}}}.
\end{eqnarray}

Now we return to the actual long chain, which is made of $N$ large granules and $N-1$ small ones.
Our effective chain then has $N$ effective granules where each of the two edge granules
corresponds to either the ``left" or the ``right" effective granule
described above, and the rest of the granules in the chain are described according to
the ``between" granule prescription. We emphasize that since the size $R_k$ of the smaller
granules is a random variable, the renormalized masses, Eq. (\ref{renorm-mass}), and the effective
interactions, Eq. (\ref{effec-int}), in the effective chain are also random variables. This
completes the mapping onto an effective chain. This mapping in turn allows us to implement the
binary collision approximation, which we do next.

\subsection{Binary collision approximation}
\label{binary}

Using the binary collision approximation, in our previous studies we successfully calculated the
time that it takes the pulse to reach the end of the chain. For this purpose we first calculate
the time spent by the pulse 
at each granule.  If we assume that in the effective chain the pulse propagates through a series of successive
binary collisions, the time taken by the pulse to go from
the $k$th granule to the $(k+1)$st granule in the chain is given by
\begin{eqnarray}
\label{tk-1}
T_{k,k+1} = A(n) \left[\frac{{\cal M}_{k,k+1}}{r_k}\right]^{1/n} v_{k}^{\frac{2}{n}-1},
\end{eqnarray}
where 
\begin{eqnarray}
\label{an}
A(n) = \sqrt{\pi}\frac{\Gamma(1+1/n)}{\Gamma(1/2+1/n)}(2^{n-2}n)^{1/n},
\end{eqnarray}
and ${\cal M}_{k,k+1}=\mu_k\mu_{k+1}/(\mu_k+\mu_{k+1})$ is the reduced mass for the pair of granules
$k$ and $(k+1)$~\cite{PRE2}. The velocity amplitude $v_k$ in Eq.~(\ref{tk-1}) is 
\begin{eqnarray}
 \label{vk}
v_k = \prod_{n=1}^{k-1} \frac{2 {\cal M}_{n,n+1}}{\mu_{n+1}}.
\end{eqnarray}
The time taken by the pulse to reach the end of the chain is therefore
\begin{eqnarray}
t_N &=& \sum_{k=1}^{N-1} T_{k,k+1}\nonumber \\
&=& A(n) \sum_{k=1}^{N-1} \left[\frac{{\cal M}_{k,k+1}}{r_k}\right]^{1/n} v_{k}^{\frac{2}{n}-1}.
\label{total-time}
\end{eqnarray}

Suppose now that the small granules are chosen randomly from a given distribution, i.e., their
sizes and consequently their masses are random variables. This means that each $k$-dependent term on the right hand side
of Eq.~(\ref{tk-1}) is a random variable. To help us manage these contributions, we implement two further approximations
for Eq.~(\ref{total-time}).
First, since the radius of the small granules is restricted to be no larger than 40\% of that of the
large granules, $R_k\leq 0.4$, we neglect the correction of order $R_k^3$ to the masses of the
(large) granules in the effective chain. All granules
in the effective chain then have equal (fixed) masses equal to that of the large masses in the
original chain [see Eq.(\ref{renorm-mass})].
Thus $v_k= 1$ and ${\cal M}_{k,k+1} = m/2$. Equation (\ref{total-time}) then reduces to
\begin{eqnarray}
\label{total-time-1}
t_N =  A(n) \left(\frac{m}{2}\right)^{1/n}\sum_{k=1}^{N-1}
\left(\frac{1+R_k}{2R_k}\right)^{\frac{1}{2n}}.
\end{eqnarray}
\begin{figure}[h]
\centering
\rotatebox{0}{\scalebox{.30}{\includegraphics{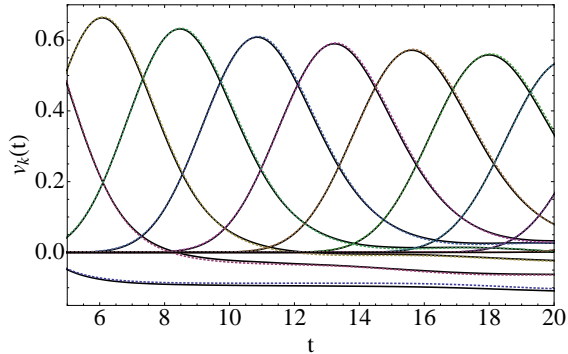}}}
\caption{(Color online) Averaged velocity profile of the larger granules with $N=11$. Results from the effective
scheme are shown as black continuous curves.}
\label{figure1}
\end{figure}
Secondly, we assume that $R_k$ is sufficiently small to neglect it in the numerator of
the above expression, which is consistent with the requirement for the validity of the effective
description in the first place. We thus approximate Eq.~(\ref{total-time-1}) as
\begin{eqnarray}
\label{total-time-2}
t_N \approx  A(n) \left(\frac{m}{2}\right)^{1/n}\sum_{k=1}^{N-1} \left(2R_k\right)^{-\frac{1}{2n}}
\end{eqnarray}
(this assumption might be questionable when $R_k$ is as large as $0.4$, but the results shown later
support the applicability of this approximation). 

Although all the formulas given above are valid for any distribution function subject to
the stated restrictions, for concreteness we implement a uniform distribution of small radii
over the interval $[\alpha,\beta]$.  It can easily be shown that each term $X_k=(R_k)^{-1/2n}$ on
the right hand side of Eq.~(\ref{total-time-2}) is then distributed
over the range $[\beta^{-\frac{1}{2n}},\alpha^{-\frac{1}{2n}}]$ according to 
\begin{eqnarray}
\label{dist-1}
P(X_k) = \frac{2n}{\beta-\alpha} X_k^{-(2n+1)}.
\end{eqnarray}
Since this distribution has finite first and second moments, as $N\to \infty$
it follows from the central limit theorem that
the sum over the independent random variables $X_k$ tends to a Gaussian distribution.
The mean and the variance of the Gaussian are obtained from the mean ($\kappa$) and the variance ($\sigma^2$)
of the underlying distribution. Thus for large $N$, $t_N$ is distributed around the mean $N\kappa$
with a variance given by $N\sigma^2$. 
It is straightforward to see that the average time $\langle t_N\rangle$ then varies linearly with $N$,
$\langle t_N\rangle \approx S(N-1)$, and the slope of the line is given by
\begin{eqnarray}
 \label{slope}
S=\frac{2nA(n)}{2n-1}\left(\frac{m}{2^{3/2}}\right)^{1/n}\left(\frac{\beta^{1-1/(2n)}
-\alpha^{1-1/(2n)}}{\beta-\alpha}\right).
\end{eqnarray}
We stress that Eq.~(\ref{total-time-2}) and consequently (\ref{slope}) are valid when the effect of
randomness is negligible on the masses of the large granules and is only important in modifying the
effective interactions between them.  Note that
within the binary collision approximation these expressions can be used not only 
to calculate the time that the pulse takes to arrive at the end of the chain but to arrive
at any specified granule of the chain.

In the next section we shall compare these results with the numerical solution of the exact
equations and also with the numerical solution of the equations for the effective chain.

\section{Comparison with numerical results}
\label{numerics}

To restate our scenario, we consider a chain of $N$ identical large granules, each pair of which
is separated by a small granule. In all our chains the first and the last granules are of the large
variety.  The radii of the $(N-1)$ small granules (or masses) are randomly selected from a
uniform distribution in the range $[\alpha,\beta]$. For all of our numerical results we
choose $\alpha=0.01$ and $\beta=0.4$. The value of $\beta=0.4$ (and no larger) is dictated by
the validity of the effective description. Note that for $\beta=0.4$, the mass of granules in
the effective chain (Eq.~\ref{renorm-mass}) can randomly increase by
up to $\sim 6\%$ of the mass of the large granules in the original chain.
The density of all granules is the same. The equation of motion for each granule is given by
Eq.~(\ref{eq:motion_rescaled}). We solve the set of $2N-1$ coupled equations numerically for a large
number of realizations ($\sim 30000$) of the radii of the small granules.
In this section we compare these solutions with the corresponding results for the effective chain,
and also with those obtained from the binary collision approximation to the effective chain.

In Fig.~\ref{figure1} we show
the velocity amplitude profile of the bigger granules averaged over all realizations.
We observe that a well-behaved
(average) pulse of time-varying amplitude and width propagates through the random chain. 
In the same figure we show the results obtained from the effective description,
where the granular masses and their interactions are given by 
Eqs.~(\ref{renorm-mass}) and (\ref{effec-int}), respectively. The two results are almost identical,
thus showing that a description in terms of the effective chain is valid even in the presence
of randomness.
\begin{figure}[h]
\centering
\rotatebox{0}{\scalebox{.25}{\includegraphics{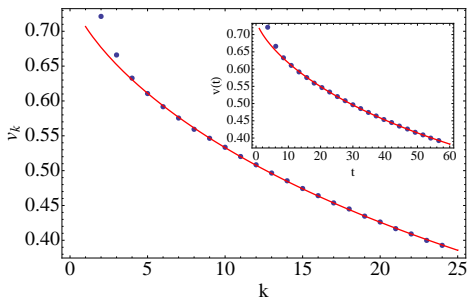}}}
\caption{(Color online) Decay in the average pulse velocity amplitude as a function of the
(larger) granular number $k$ in the effective chain
with $N=26$.  The continuous curve represents the
stretched exponential decay $v_k=a\exp(-bk^c)$ with $a=0.756, b=0.067$ and $c=0.714$. Inset:
stretched exponential decay in the pulse velocity amplitude as a function of time $t$.}
\label{figure2}
\end{figure}

In Fig.~\ref{figure2} we show the change in the pulse amplitude as it passes from one large granule to the
next along the chain. The results are indistinguishable between the original decorated chain and
the effective chain.  The amplitude follows a stretched exponential decay $v_k=a\exp(-bk^c)$ with
fitted parameter values $a=0.756$, $b=0.067$, and $c=0.714$. The inset in the figure shows the stretched
exponential decay in the pulse amplitude as a function of time, 
$v(t)=a^\prime\exp(-b^\prime t^{c^\prime})$ with $a^\prime = 0.744$, $b^\prime = 0.035$, and
$c^\prime= 0.714$. Note that $c=c^\prime$, indicating that time $t$ and granule number $k$ are
linearly related, i.e., $t\propto k$.

\begin{figure}[h]
\centering
\rotatebox{0}{\scalebox{.25}{\includegraphics{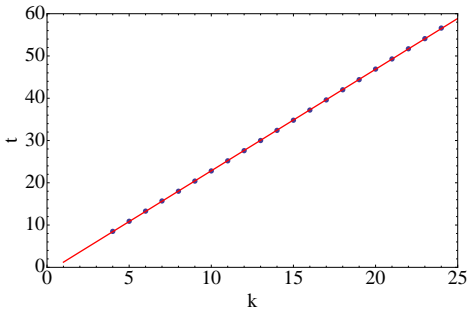}}}
\caption{(Color online) Average pulse propagation time as a function of $k$ for $N=26$ for the 
effective chain. The straight line is the fit $t= 2.402 k-1.181$.}
\label{figure3}
\end{figure}

The variation of $t$ with $k$ in the original chain or in the effective chain (they are again
indistinguishable) is shown in Fig.~\ref{figure3} for $N=26$. The
linear relation between $k$ and $t$ was obtained in our earlier work on pulse propagation in
monodisperse granular chains~\cite{alexandremono,alexandre}.
The slope of this linear variation in time as obtained from the fit to the numerical data
is $\approx 2.402$.  This is within $2\%$ of the value $S= 2.348$ obtained from the binary
colision approximation, Eq.~(\ref{slope}). Thus the effective chain as well as the binary collision
approximation to it yield excellent results in agreement with those of the original chain for the
variation of the pulse propagation time as a function of granule number. 

\begin{figure}[h]
\vspace{.6cm}
\centering
$\begin{array}{cc}
\rotatebox{0}{\scalebox{.20}{\includegraphics{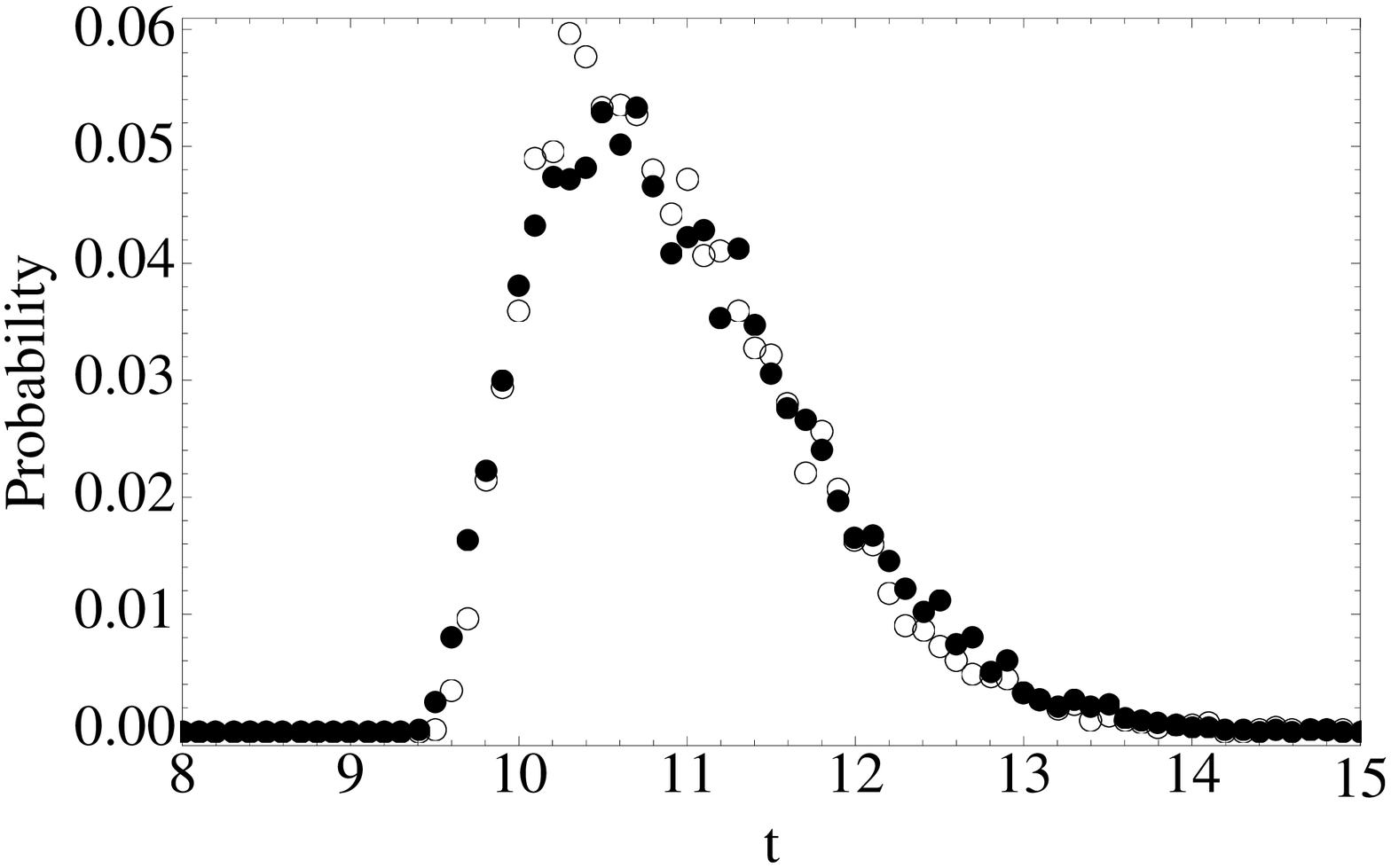}}} &
\rotatebox{0}{\scalebox{.20}{\includegraphics{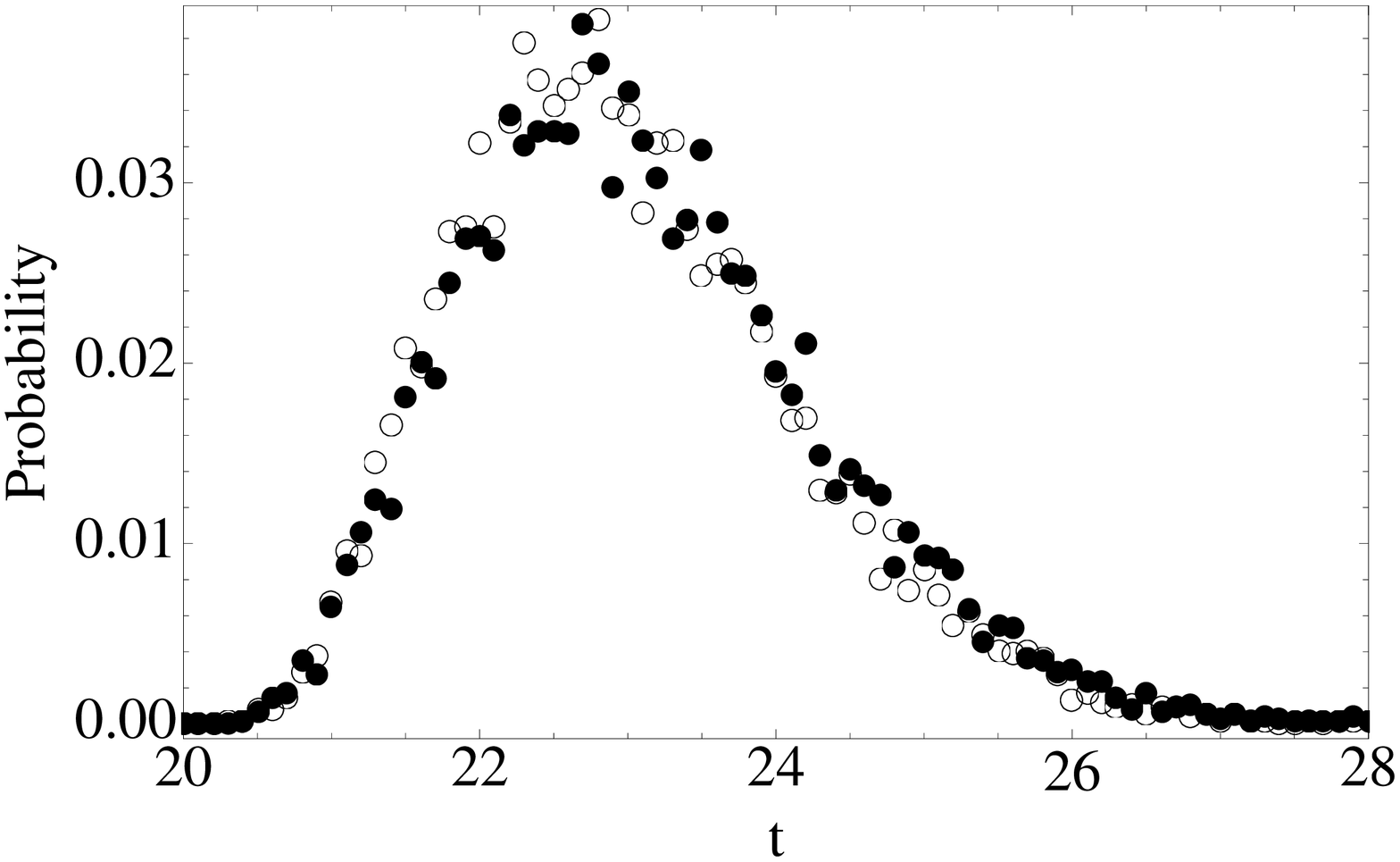}}}\\
\rotatebox{0}{\scalebox{.20}{\includegraphics{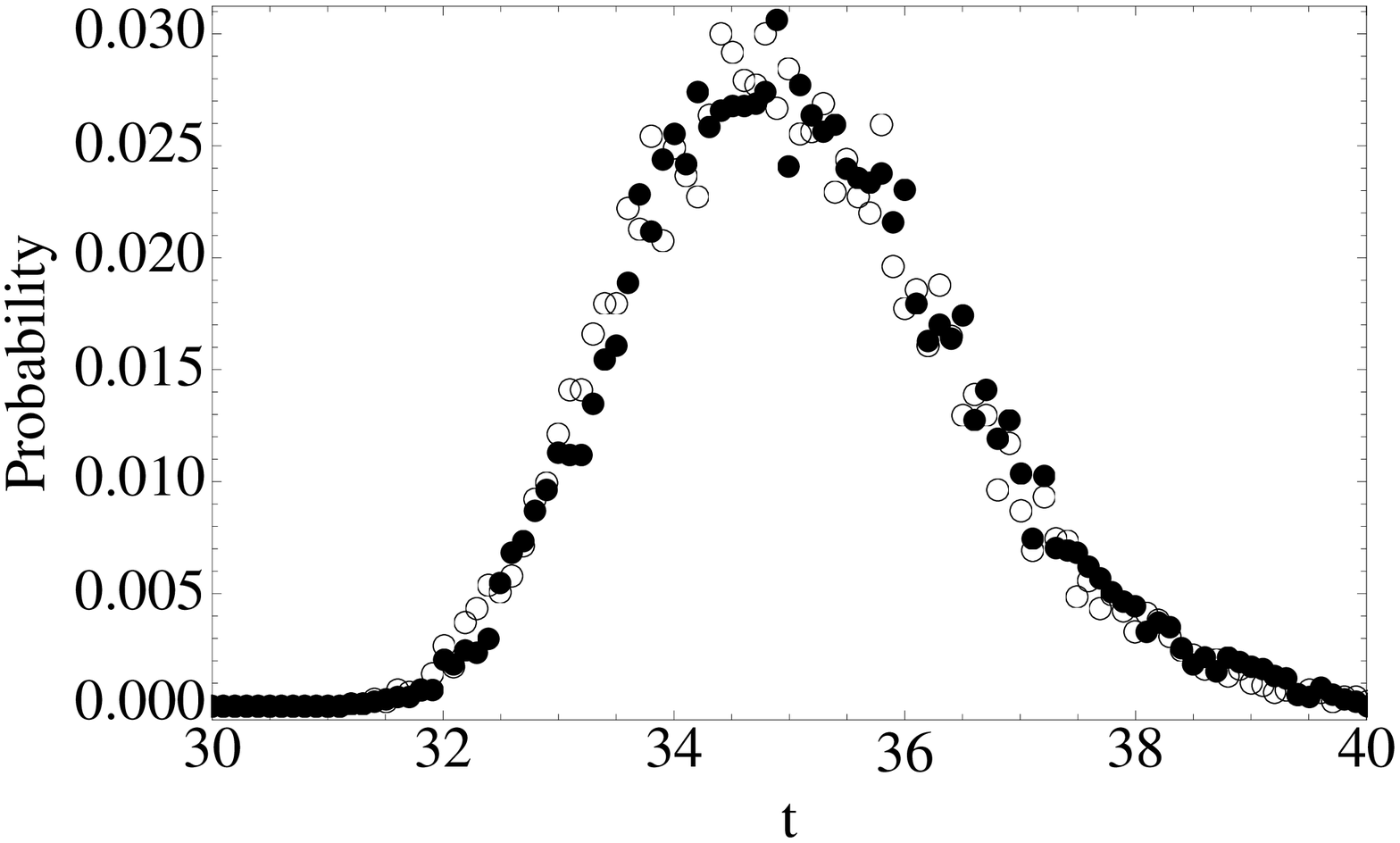}}} &
\rotatebox{0}{\scalebox{.20}{\includegraphics{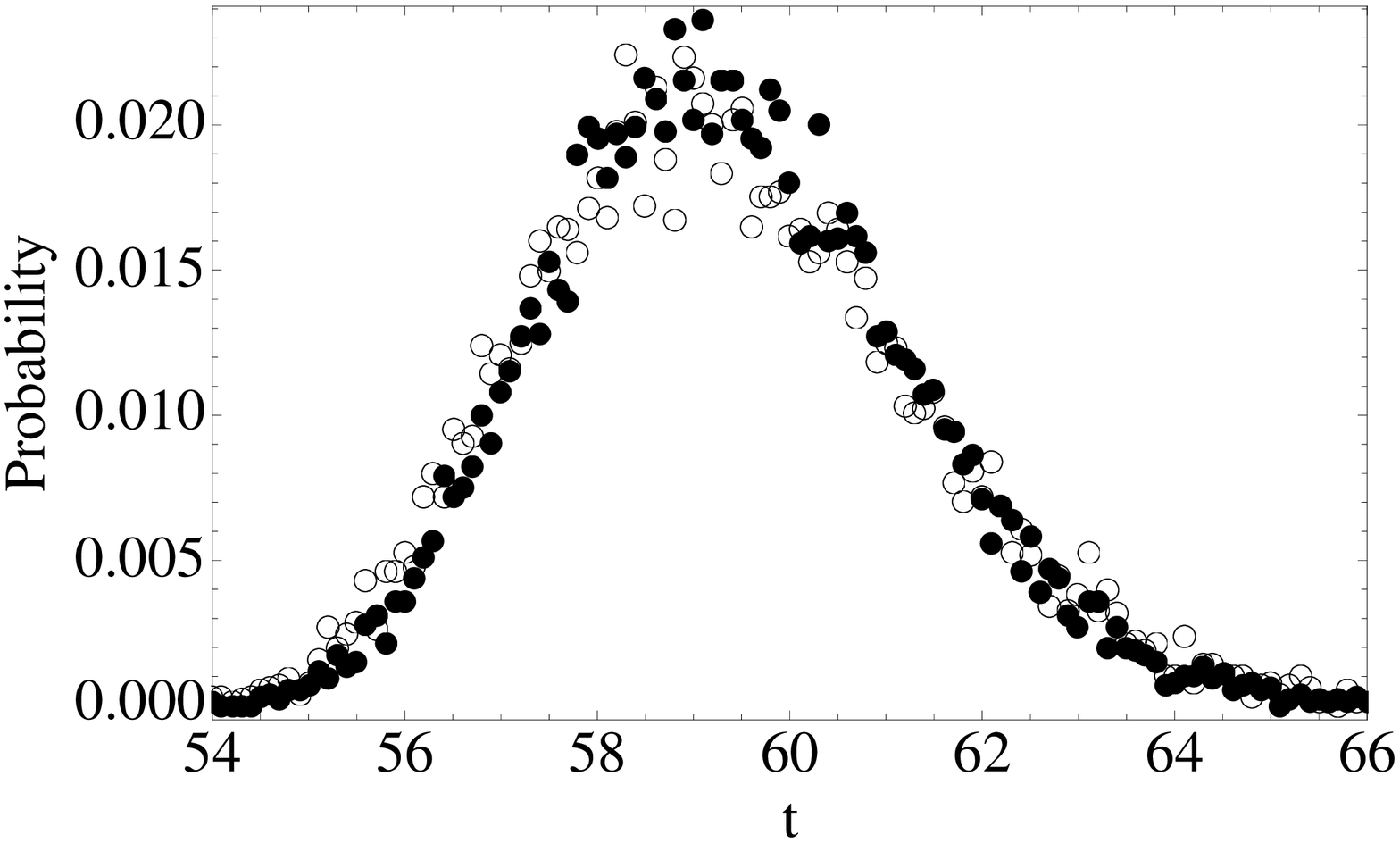}}}
\end{array}$
\caption{The distribution of arrival times of the pulse to granule $k=2N-3$.
This is the large granule before the last large granule at the end of a chain of $2N-1$ granules.
Here $N=6,11,16,26$ starting from the left upper panel and changing in the clockwise direction.
Filled circles: theory; empty circles: numerics.}
\label{figure4}
\end{figure}

The effective chain captures the behavior of the original chain extremely well for all the
properties considered above.  Indeed, the behavior of the two is essentially indistinguishable.  We
now go on to assess the validity of the binary collision approximation for the effective chain.  Above we
showed that the average pulse propagation time as a function of granule number is captured very accurately
by the approximation.
Next we focus on the distribution of times that the pulse takes to reach the end of the chain.
Because of the randomness in the chain, this time is distributed around a mean. We compute
the distribution of this time by solving the exact dynamical
equations of motion for the decorated chain
over a large number of realizations ($\sim$ 30000) of the distribution of smaller granules.
The theoretical prediction
for this time obtained from the binary collision model 
is given by Eq.~(\ref{total-time}).
In Fig.~\ref{figure4} we show comparisons between the theory (filled circles) and 
the numerical results (empty circles) for various lengths of the chain. In showing the
comparison in Fig.~\ref{figure4},
we have adjusted the peak position of the 
distribution obtained from the theory, which gives slightly higher values of the peak position.
This shift arises from the small error in the prediction of the pulse velocity when using the binary
collision approximation.  In calculating the distribution of arrival times, this small error is
accumulated over all the terms in the sum, that is, it is in effect multiplied by $N$,
the length of the chain.  However, we note that the amplitude and the shape of
the distribution are very well reproduced by the theory.

For a short chain ($N=6,11$) the distribution of arrival times of the pulse at the
edge of the chain is quite asymmetric. This asymmetry decreases as the length of the chain
is increased, and for $N=26$ the distribution is approaching a Gaussian. This is the result of the
central limit theorem.  Remember that in Eq.~(\ref{total-time}) we are adding random 
terms. However these terms are {\em not} independent (each term contains information about the random
size/mass of all the previous granules through $v_k$),
and therefore the sum is more complicated than one involving independent random variables.
In any case, since for large $N$ the sum assumes a Gaussian-like form, the correlation between
different terms in Eq.~(\ref{total-time}) is presumably small or highly localized (falling off quickly
with increasing $k$). 

In order to quantify the difference between the computed distribution functions in Fig.~\ref{figure4}
and the standard normal distribution, we have performed the Kolomogorov-Smirnov~\cite{KS}
statistical test on the data obtained numerically and used to generate the distribution
shown in the figure.  For this purpose, we first collect the data $\{X_n=x_1,x_2,\cdots x_n\}$ computed
from the numerical solution in increasing order $x_1<x_2<\cdots x_n$. Then the mean $\mu_n$ and the standard
deviation $\sigma_n$ are computed as
\begin{eqnarray}
 \label{AD}
\mu(n) &=& \frac{1}{n}\sum_{i=1}^n x_i\nonumber\\
\sigma(n) &=& \sqrt{\frac{1}{n-1}\sum_{i=1}^n (x_i-\mu)^2}.
\end{eqnarray}
A new data set $z_n=(X_n-\mu(n))/\sigma(n)$ is then generated. The Kolmogorov-Smirnov test involves computing
the statistics of the absolute difference (non-directional hypothesis) between the cumulative
frequency distribution 
$F_z(n)$ of $z_n$ and that of the standard normal distribution $F_0$, i.e.,
$d=|F_z(n)-F_0|$~\cite{KS}. Acceptance of the null hypothesis that the distribution is Gaussian
within a given level of confidence requires that $d$ be appropriately small. Consistent with the
progression seen in Fig.~\ref{figure4}, we find a steady decrease in the value of $d$ as $N$
increases, that is, our distribution approaches a Gaussian.

Finally, we turn to the velocity amplitude, which while extremely well reproduced in the effective
undecorated chain is not captured well by the binary collision approximation. 
It is easy to see the source of the problem.
Recall that Eq.~(\ref{slope}) is valid only when the masses of the
granules in the effective chain are assumed to all be equal and only the interactions between them
are affected by the randomness of the sizes of the
smaller granules. The agreement between the theory, Eq.~(\ref{slope}), and the numerics indicates that
these assumptions are valid when calculating pulse travel times. However,
the velocity obtained from the model as posed in Eq.~(\ref{vk})
is independent of $k$ (and only very weakly dependent on $k$ if the direct $\lesssim6\%$ effect of the randomness
of the masses is not neglected) and does not follow the behavior
observed in the numerical solution of the exact equations, Fig.~\ref{figure2}.
The problem lies in the fact that the velocity $v_k$ obtained from the binary model depends only
on the mass ratio of two colliding granules.  However, the random interactions can introduce
unavoidable three (or more) granule scenarios.  For example, if the interaction between a granule
$1$ and a granule $2$ is weak but that between $2$ and $3$ is strong, then a collision between $1$
and $2$ will inevitably involve granule $3$.  A remedy might be to include this effect through a
further renormalization of the masses resulting from the random interactions.  This will be
explored in future work.

\section{conclusions}
\label{conclusion}

We have studied pulse propagation in 1D granular chains decorated with small granules of random
radii inserted between each pair of large granules all of the same size.
This study has proceded in two steps.

Firstly, we used the effective scheme introduced in
Ref.~\cite{PRE2} to obtain an equivalent undecorated
random chain, and showed that this effective description works
remarkably well for all properties tested, so well that the numerical results obtained from the
original chain and from the effective chain are essentially indistinguishable.  Since the effective
chain is only half as long as the original chain, this represents a considerable savings in
computational effort.

Secondly, using the binary collision approximation on the
effective chain, we have obtained analytic expressions 
for the velocity amplitude of the pulse and the time that the 
pulse takes to reach the $k$th granule (pulse speed $dk/dt$) along the chain.
By construction, the analytic results obtained using the binary collision approximation neglect the
effects of the randomness of the small granules on the masses of the large granules in the effective
chain.  The randomness appears only in the interactions between granules.
It is thus not surprising that the velocity amplitude of the pulse obtained from the binary model 
does not predict the correct behavior as seen in numerical results, since this amplitude is
determined by the masses of colliding granules and is therefore affected 
by the randomness of these masses. In the previous section we have suggested a possible remedy to
this issue.  On the other hand,
the pulse speed and the distribution of the times taken by the pulse to reach the $k$th
granule are very well reproduced by the theory. This is because
the time of pulse propagation depends on the interactions, being shorter (longer) for
stronger (weaker) interactions. Our theory incorporates this dependence very accurately,
cf. Eq.~(\ref{total-time}).

As was noted in Ref.~\cite{PRE2},
the effective description works well as long as the size of the small granule remains
less than $\sim 40\%$ of the bigger granule. This places a restriction on the size/mass
distribution of the smaller granules. Here, for simplicity, we have considered a
uniform distribution of the smaller granules. However, the validity of the effective description
and of the binary-collision approximation will remain valid for arbitrary distributions as
long as the size restriction on the small granules is satisfied.  In future work we plan to
generalize our effective chain description (and the associated binary collision approximation) to
other chain configurations, with the eventual goal of understanding pulse propagation in chains of
arbitrary granular configurations.

\section*{Acknowledgments}
Acknowledgment is made to the Donors of the American Chemical Society Petroleum Research Fund for
partial support of this research (K.L. and U.H.).  A.R. acknowledges support from Bionanotec-CAPES
and CNPq.  A. H. R. acknowledges support by CONACyT Mexico under Projects J-59853-F and J-83247-F.

\end{document}